\documentclass[epj]{svjour}
\usepackage[dvips]{graphicx}
\usepackage{amsmath}
\usepackage{amssymb}
\usepackage{subeqnarray}
\usepackage{subfigure}

\begin{document}

\title{Detuning effects in the vertical cold-atom micromaser}
\author{John Martin\inst{1}\thanks{email: jmartin@irsamc.ups-tlse.fr} \and Thierry Bastin\inst{2}\thanks{email: T.Bastin@ulg.ac.be}}

\institute{Laboratoire de Physique Th\'eorique, Universit\'e de
Toulouse III, CNRS, 31062 Toulouse France \and Institut de Physique
Nucl\'eaire, Atomique et de Spectroscopie, Universit\'e de Li\`ege,
B\^at.\ B15, B - 4000 Li\`ege, Belgium}

\date{June 12, 2008}

\abstract{The quantum theory of the cold atom micromaser including
the effects of gravity is established in the general case where the
cavity mode and the atomic transition frequencies are detuned. We
show that atoms which classically would not reach the interaction
region are able to emit a photon inside the cavity. The system turns
out to be extremely sensitive to the detuning and in particular to
its sign. A method to solve the equations of motion for non resonant
atom-field interaction and arbitrary cavity modes is presented.
 } \maketitle

\section{Introduction}
\label{Introduction}

\begin{figure}
\begin{center}
\includegraphics[width=.95\linewidth]{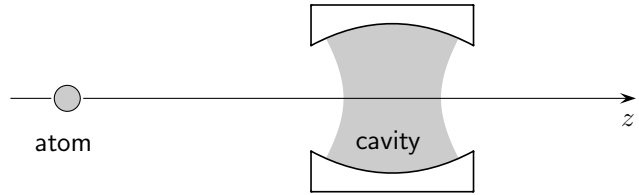}
\end{center}
\caption{Scheme of the horizontal micromaser.} \label{hm}
\end{figure}

The coupling of atomic motion to light is a topic of major interest
in quantum optics. The mechanical effects of (laser) light on atoms
can be exploited to cool, trap and handle the atoms with a great
accuracy~\cite{Coh98} (see for example~\cite{Mir06} for manipulation
of individual atoms with optical tweezers). The achieved control on
the atomic motion gives rise to a host of applications, like the
development of new optical frequency standards based on single
ions~\cite{Kle00}, the realization of an atom laser~\cite{Hag99},
the possibility of implementing electronic components with atoms
(instead of electrons)~\cite{Rus04,Rus06,Sea07}, or quantum
information processing with cold atoms and trapped
ions~\cite{Gar05}. In the last example, the interplay between
external (motional) and internal degrees of freedom plays a
prominent role. This is also the case in the usual micromaser (see
Fig.~\ref{hm}) when cold atoms rather than thermal ones are sent
through the cavity. Indeed, in this regime a new type of induced
emission has been shown to occur because the wave behavior of the
atoms becomes important~\cite{Scu96}. The resulting process called
microwave amplification via $z$-motion-induced emission of radiation
(\emph{mazer})~\cite{Scu96} has opened up a new chapter of
micromaser
physics~\cite{Mey97,Lof97,Lof98,Aga00,Aru00,Aru02,Bas03b,Mar04,Sei07,Mar07}.
In all these papers, gravity effects on the quantized atomic motion
are not considered. However, it has often been argued that the
achievement of the cold atom micromaser would be a formidable
experimental task as the atoms in this regime move so slowly that
they start exactly to be extremely sensitive to earth gravity and
are expected to fall down before entering or leaving the cavity
region, breaking the unidirectionality of the atomic
trajectories~\cite{Wal06}. A possible way to deal with gravity for
the mazer action while keeping the atomic trajectories
one-dimensional is to consider a vertical geometry where the atoms
are sent vertically in the direction of the cavity, similarly to
cavity QED experiments reported
in~\cite{Mab96,Hoo98,Mun99a,Mun99b,Ye99,Hen00,Shi02,Mau04}.
Recently, we established the quantum theory of such a vertical
mazer, taking into account gravity effects on the vertical quantized
atomic motion~\cite{Bas05a}. The theory was written for two-level
atoms in the resonant case where the cavity mode frequency $\omega$
is equal to the atomic transition frequency $\omega_0$. Here, we
remove this restriction and investigate the effects of a detuning on
the quantum evolution of the combined atom-cavity system in a
vertical configuration where gravity action is taken into account.

The paper is organized as follows. In Sec.~\ref{ModelSection}, the
Hamiltonian modeling our system is presented. The properties of the
induced emission probability of a photon inside the cavity are then
presented in Sec.~\ref{PemSection}. The connection with the
classical regime is discussed. A brief summary of our results is
finally given in Sec.~\ref{SummarySection}. In Appendix A, a
technical derivation of a specific formula used in our paper is
derived. The method we have developed to solve numerically the
equations of motion of the mazer in the non resonant case for any
mode function is given in Appendix B.

\section{Model}
\label{ModelSection}

\subsection{The Hamiltonian}

\begin{figure}
\begin{center}
\includegraphics[width=.45\linewidth]{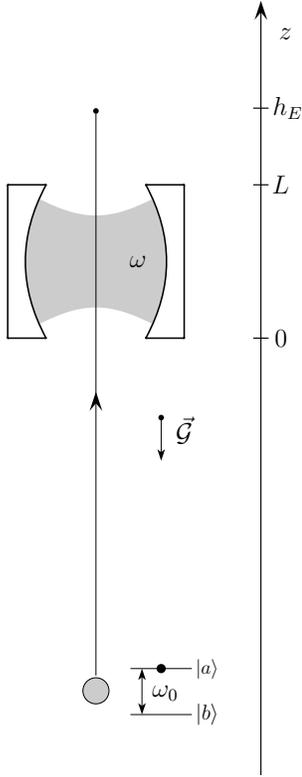}
\end{center}
\caption{Scheme of the vertical mazer.} \label{vm}
\end{figure}

We consider a two-level atom moving in the gravity field along the vertical
$z$ direction on the way to a cavity which we define to be located
in the range $0<z<L$ (see Fig.~\ref{vm}). No transverse motion is considered.
The atom is coupled nonresonantly to a single mode of the quantized
field present in the cavity. The atomic center-of-mass motion is
described quantum mechanically and the usual rotating-wave
approximation is made. The Hamiltonian thus reads
\begin{equation}
    \label{Hamiltonian}
        H = \hbar \omega_0 \sigma^{\dagger} \sigma + \hbar \omega a^{\dagger} a + \frac{p^2}{2m}+ m\mathcal{G} z+ \hbar g \, u(z) (a^{\dagger} \sigma + a
        \sigma^{\dagger}),
\end{equation}
where $p$ is the atomic center-of-mass momentum along the $z$
axis, $m$ is the atomic mass, $\mathcal{G}$ is the acceleration
due to gravity, $\omega_0$ is the atomic transition frequency,
$\omega$ is the cavity field mode frequency, $\sigma = |b \rangle
\langle a|$ ($|a\rangle$ and $|b\rangle$ are, respectively, the
upper and lower levels of the two-level atom), $a$ and
$a^{\dagger}$ are, respectively, the annihilation and creation
operators of the cavity radiation field, $g$ is the atom-field
coupling strength, and $u(z)$ the cavity field mode function. We
denote in the following the detuning $\omega-\omega_0$ by
$\delta$, the cavity field eigenstates by $|n\rangle$, the global
state of the atom-field system at any time $t$ by $|\psi(t)\rangle$, and the
classical height attained by an atom of energy $E$ in the gravity
field by
\begin{equation}\label{he}
h_E\equiv\frac{E}{m\mathcal{G}}.
\end{equation}

\subsection{The wave functions}\label{wfunctions}
We introduce the orthonormal basis
\begin{equation}
\begin{aligned}
\label{basis} &|\Gamma_{n}^+(\theta)\rangle =
\cos\theta\,|a,n\rangle +
\sin\theta\,|b,n+1\rangle,\\
&|\Gamma_{n}^-(\theta)\rangle = -\sin\theta\,|a,n\rangle +
\cos\theta\,|b,n+1\rangle,
\end{aligned}
\end{equation}
with $\theta$ an arbitrary parameter. The
$|\Gamma_{n}^{\pm}(\theta)\rangle$ states coincide with the
noncoupled states $|a,n\rangle$ and $|b,n+1\rangle$ when $\theta =
0$ and with the dressed states when $\theta = \theta_n$ given by
\begin{equation}
    \label{thetads}
    \cot 2 \theta_n = - \frac{\delta}{\Omega_n}\, ,
\end{equation}
with the Rabi frequency
\begin{equation}
    \Omega_n = 2 g \sqrt{n + 1}.
\end{equation}

We denote as $|\pm, n\rangle$ the dressed states
$|\Gamma_{n}^{\pm}(\theta_n)\rangle$. The Schr\"odinger equation
reads in the $z$ representation and in the basis (\ref{basis})
\begin{subequations}
\label{ss}
\begin{equation}
\begin{aligned}
\label{ss+} & i\hbar\frac{\partial}{\partial t}\psi^+_{n,
\theta}(z,t)\; = \;\bigg[
  -\frac{\hbar^2}{2m}\frac{\partial^2}{\partial
    z^2}+m\mathcal{G}z+(n+1)\hbar\omega\\&  \;\;-\cos^2\theta\,
    \hbar\delta
    +\hbar g\,
  u(z) \sqrt{n\! +\! 1}\:\sin2\theta\bigg]\psi^+_{n, \theta}(z,t) \\&
+\bigg[\hbar g u(z) \sqrt{n\! +\!
1}\:\cos2\theta+\frac{\sin2\theta}{2}\:\hbar\delta\bigg]\psi^-_{n,
\theta}(z,t),
\end{aligned}
\end{equation} \vspace{-0.2cm}
\begin{equation}
\begin{aligned}\label{ss-}
& i\hbar\frac{\partial}{\partial t}\psi^-_{n, \theta}(z,t)\; =
\;\bigg[
  -\frac{\hbar^2}{2m}\frac{\partial^2}{\partial
    z^2}+m\mathcal{G}z+(n+1)\hbar\omega\\ & \;\;-\sin^2\theta\:\hbar\delta-\;\hbar g
  u(z) \sqrt{n\! +\! 1}\:\sin2\theta\bigg]\psi^-_{n, \theta}(z,t)\\ &
+\bigg[\hbar g u(z) \sqrt{n\! +\!
1}\:\cos2\theta+\frac{\sin2\theta}{2}\:\hbar\delta\bigg]\psi^+_{n,
\theta}(z,t),
\end{aligned}
\end{equation}
\end{subequations}
with
\begin{equation}
    \psi^{\pm}_{n, \theta}(z,t) = \langle z,
    \Gamma_{n}^{\pm}(\theta)| \psi(t)\rangle.
\end{equation}

We get for each $n$ two coupled partial differential equations. In
the resonant case ($\delta = 0$), these equations may be decoupled
over the entire $z$ axis when working in the dressed state basis and
the atom-field interaction reduces to an elementary scattering
problem in the presence of the gravitational field over a potential
barrier and a potential well defined by the cavity~\cite{Bas05a}. In
the presence of a detuning, this is no longer the case~: there is no
basis where Eqs.~(\ref{ss}) would separate over the entire $z$ axis
and the interpretation of the atomic interaction with the cavity as
a scattering problem over two potentials is less obvious as it is
also the case when gravity is not taken into account~\cite{Bas03b}.

In the noncoupled state basis ($\theta = 0$), Eqs.~(\ref{ss}) read
\begin{subequations}
\label{ssab}
\begin{equation}
\begin{aligned}
\label{ssa}  i\hbar\frac{\partial}{\partial t}\psi^a_n(z,t) =
&\left[
  -\frac{\hbar^2}{2m}\frac{\partial^2}{\partial
    z^2}+m\mathcal{G}z \right]\psi^a_n(z,t) \\&
+\hbar g u(z) \sqrt{n\! +\! 1}\:\psi^b_{n+1}(z,t),
\end{aligned}
\end{equation} \vspace{-0.2cm}
\begin{equation}
\begin{aligned}\label{ssb}
i\hbar\frac{\partial}{\partial t}\psi^b_{n + 1}(z,t) = &  \left[
  -\frac{\hbar^2}{2m}\frac{\partial^2}{\partial
    z^2}+m\mathcal{G}z + \hbar \delta \right]\psi^b_{n +
    1}(z,t)\\ &
+\hbar g u(z) \sqrt{n\! +\! 1}\:\psi^a_{n}(z,t),
\end{aligned}
\end{equation}
\end{subequations}
with
\begin{subequations}
\label{psiapsib}
\begin{eqnarray}
\psi^a_n(z,t) & = & e^{i(\omega_0 + n \omega) t}\, \langle z,a,n|\psi(t)\rangle, \\
\psi^b_{n + 1}(z,t) & = & e^{i(\omega_0 + n \omega) t}\, \langle
z, b,n+1|\psi(t)\rangle.
\end{eqnarray}
\end{subequations}

In Eqs.~(\ref{psiapsib}), we have introduced the exponential factor
$e^{i(\omega_0 + n \omega) t}$ in order to define the energy scale
origin at the $|a,n\rangle$ level. If we assume initially a
monoenergetic excited atom coming upwards upon the cavity that
contains $n$ photons, the atom-field system is described outside the
cavity by the wave function components (which correspond to the
eigenstate $|\phi_E\rangle$ of energy $E$)
\begin{subequations}
\label{psiapsibn}
\begin{eqnarray}
\psi^a_n(z,t)  & = & e^{-iEt/\hbar}\, \varphi^a_{E,n}(z), \\
\psi^b_{n+1}(z,t)  & = & e^{-iEt/\hbar}\, \varphi^b_{E,n+1}(z),
\end{eqnarray}
\end{subequations}
with $\varphi^a_{E,n}(z)$ and $\varphi^b_{E,n+1}(z)$ obeying
\begin{subequations}
\label{ssati}
\begin{align}
&\left(-\frac{\hbar^2}{2m}\frac{\mathrm{d}^2}{\mathrm{d}
    z^2}+m\mathcal{G}z-E \right)\varphi^a_{E,n}(z)=\nonumber\\
    &\hspace{85pt}-\hbar g u(z) \sqrt{n\! +\! 1}\:\varphi^b_{E,n+1}(z), \label{ssatia}\\
&\left(-\frac{\hbar^2}{2m}\frac{\mathrm{d}^2}{\mathrm{d}
    z^2}+m\mathcal{G}z -E + \hbar \delta \right)\varphi^b_{E,n +
    1}(z)=\nonumber\\
    &\hspace{85pt}-\hbar g u(z) \sqrt{n\! +\! 1}\:\varphi^a_{E,n}(z).\label{ssatib}
\end{align}
\end{subequations}

Introducing, for any length $x$, the dimensionless variable
\begin{equation}
\label{dimensionlessx} \tilde{x}\equiv x/\ell
\end{equation}
with
\begin{equation}\label{ell}
\ell =\left(\frac{2m^2\mathcal{G}}{\hbar^2}\right)^{-1/3},
\end{equation}
Eqs.~(\ref{ssati}) read
\begin{subequations}\label{sysreduitmv}
\begin{align}
        & \left(\frac{\mathrm{d}^2}{\mathrm{d}
        \tilde{z}^2}- \tilde{z}+ \tilde{h}_E\right)\varphi^{a}_{E,n}(\tilde{z}) = \tilde{h}_{\mathrm{int}}
        \,u(\tilde{z})\,\varphi^{b}_{E,n+1}(\tilde{z}),\label{sysreduitmva}\\
        & \left(\frac{\mathrm{d}^2}{\mathrm{d}
        \tilde{z}^2}-\tilde{z} + \tilde{h}_E - \tilde{\delta}\right)\varphi^{b}_{E,n+1}(\tilde{z})
        = \tilde{h}_{\mathrm{int}}
        \,u(\tilde{z})\,\varphi^{a}_{E,n}(\tilde{z}),\label{sysreduitmvb}
\end{align}
\end{subequations}
where
\begin{equation}
\tilde{h}_{\mathrm{int}}=\tilde{g}\sqrt{n+1},
\end{equation}
and
\begin{equation}
\tilde{g}=\frac{\hbar g}{m\mathcal{G}\ell},\hspace{10pt}
\tilde{\delta}=\frac{\hbar \delta}{m\mathcal{G}\ell}.
\end{equation}

The characteristic length $\ell$ (Eq.~(\ref{ell})) gives the spatial
scale of the oscillations of the wave function of a particle in the
gravitational field near its turning point. For rubidium atoms,
$\ell\simeq 0.3\,\mu$m and, roughly, the dimensionless variable
$\tilde{\delta}$ is numerically equal to the detuning $\delta/2\pi$
expressed in kHz. Similarly $\tilde{h}_E$ and $\tilde{L}$ yield the
classical height $h_E$ and the cavity length $L$, respectively, in
thirds of $\mu$m~\cite{Bas05a}.

Outside the cavity, the mode function $u(z)$ vanishes and
Eqs.~(\ref{sysreduitmv}) read
\begin{subequations}
\label{ssatirv}
\begin{align}
&\left(\frac{\mathrm{d}^2}{\mathrm{d}
    \tilde{z}^2}-\tilde{z}+\tilde{h}_E \right)\varphi^a_{E,n}(\tilde{z})=0, \\
&\left(\frac{\mathrm{d}^2}{\mathrm{d}
    \tilde{z}^2}-\tilde{z}+\tilde{h}_E - \tilde{\delta} \right)\varphi^b_{E,n +
    1}(\tilde{z})=0.
\end{align}
\end{subequations}

The solutions to Eqs.~(\ref{ssatirv}) are given by linear
combinations of Airy Ai and Bi functions~\cite{Abr70} and may be
written in the form
\begin{align}
\varphi^a_{E,n}(z) & =\left\{
\begin{array}{lr}\label{sola}
a_{E,n}^a\,\mathrm{Ai}(\tilde{z}-\tilde{h}_E) & \hspace{10pt} z>L\vspace{5pt}\\
\mathrm{U}(\tilde{z}-\tilde{h}_E)+d_{E,n}^a\,\mathrm{D}(\tilde{z}-\tilde{h}_E)
& \hspace{10pt} z<0
\end{array}\right.\\\nonumber\\
\varphi^b_{E,n+1}(z) & =\left\{
\begin{array}{lr}\label{solb}
a_{E,n+1}^b\,\mathrm{Ai}(\tilde{z}-\tilde{h}_E+\tilde{\delta}) & \hspace{10pt} z>L\vspace{5pt}\\
d_{E,n+1}^b\,\mathrm{D}(\tilde{z}-\tilde{h}_E+\tilde{\delta}) &
\hspace{10pt} z<0
\end{array}\right.
\end{align}
with
\begin{subequations}\label{updown}
\begin{align}
        \mathrm{U}\left(\tilde{z}\right)=\mathrm{Ai}\left(\tilde{z}\right)+i\,\mathrm{Bi}\left(\tilde{z}\right)\\
        \mathrm{D}\left(\tilde{z}\right)=\mathrm{Ai}\left(\tilde{z}\right)-i\,\mathrm{Bi}\left(\tilde{z}\right)
\end{align}
\end{subequations}

Equations (\ref{sola}) and (\ref{solb}) underline that the wave
functions below the cavity consist of an upward and a downward wave
($\mathrm{U}$ and $\mathrm{D}$, respectively). Indeed, the
probability current densities associated to these two waves are
respectively given by
\begin{equation}\label{current}
j = \pm \frac{\hbar}{\pi m \ell}.
\end{equation}

For large negative $\tilde{z} - \tilde{h}_E$ values, these two waves
behave similarly to plane waves with a $z$ dependent wave
vector~\cite{Bas05a}.

Consequently, the following interpretation must be given to
solutions (\ref{psiapsibn}). The excited atom coming upwards upon
the cavity will be found propagating downwards in the upper state or
in the lower state with amplitude $d_{E,n}^a$ and $d_{E,n+1}^b$,
respectively. However, in contrast to the resonant case, the atom
propagating downwards in the lower state $|b\rangle$ possesses a
total external energy $E-\hbar\delta$ different from its initial
value $E$. The atomic transition $|a\rangle \rightarrow |b\rangle$
induced by the cavity is responsible for a change of the total
external atomic energy. According to the sign of the detuning, the
cavity will either provide energy to the atom (for $\delta < 0$) or
remove energy from the atom (for $\delta> 0$). This results merely
from energy conservation. When, after leaving the cavity region, the
atom is passed from the excited state $|a\rangle$ to the lower state
$|b\rangle$, the photon number has increased by one unit in the
cavity and the internal energy of the atom-field system has varied
by the quantity $\hbar \omega - \hbar \omega_0 = \hbar \delta$. This
variation needs to be exactly counterbalanced by the external energy
of the system, \emph{i.e.}, the total external atomic energy.

Inside the cavity, the problem is much more complex since we have
two coupled partial differential equations. In the special case of
the mesa mode function [$u(z) = 1$ inside the cavity, 0
elsewhere], the problem is, however, greatly simplified. In the
dressed state basis ($\theta = \theta_n$), the Schr\"odinger
equations (\ref{ss}) take the following form inside the cavity~:
\begin{equation}
\label{seqinside} i\hbar\frac{\partial}{\partial
t}\psi^{\pm}_n(z,t) = \left[
  -\frac{\hbar^2}{2m}\frac{\partial^2}{\partial
    z^2}+m\mathcal{G}z+ V^{\pm}_n\right]\psi^{\pm}_n(z,t)
\end{equation}
with
\begin{equation}
    \label{psipm}
    \psi^{\pm}_n(z,t) = e^{i(\omega_0 + n \omega) t} \langle
    z,\pm,n|\psi(t)\rangle
\end{equation}
and
\begin{subequations}
\begin{align}
    V^+_n & = \sin^2 \theta_n \, \hbar \delta + \hbar g \sqrt{n + 1} \, \sin 2 \theta_n, \\
    V^-_n & = \hbar \delta - V^+_n. \label{VnmasVnp}
\end{align}
\end{subequations}

Using Eq.~(\ref{thetads}), we have the well known
relations~\cite{Coh88}
\begin{equation}\label{cot}
\begin{array}{lr}
\sin\theta_{n}=\frac{\sqrt{\Omega'_n+\delta}}{\sqrt{2\Omega'_n}}, &
\;\;\;\tan\theta_{n}=\sqrt{\frac{\Omega'_n+\delta}{\Omega'_n-\delta}},
\vspace{0.3cm}\\
\cos\theta_{n}=\frac{\sqrt{\Omega'_n-\delta}}{\sqrt{2\Omega'_n}},
&\;\;\;
\cot\theta_{n}=\sqrt{\frac{\Omega'_n-\delta}{\Omega'_n+\delta}},
\end{array}
\end{equation}
with the generalized Rabi frequency
\begin{equation}
\Omega'_n=\sqrt{\Omega_n^2+\delta^2}.
\end{equation}

We thus have
\begin{subequations}
\begin{align}
    V^+_n & = \hbar g \sqrt{n + 1}\, \tan \theta_n, \\
    V^-_n & = -\hbar g \sqrt{n + 1}\, \cot \theta_n.
\end{align}
\end{subequations}

The exponential factor $e^{i(\omega_0 + n \omega) t}$ has been
introduced as well in Eq.~(\ref{psipm}) in order to define the
same energy scale inside and outside the cavity. The positive
internal energy $V^+_n$ increases with positive detunings and vice
versa with negative ones. For large positive (resp.\ negative)
detunings, $V^+_n$ tends to the $|b,n+1\rangle$ (resp.\
$|a,n\rangle$) state energy.

The most general solution of Eqs.~(\ref{seqinside}) is given by
\begin{equation}
\label{psipminside} \psi^{\pm}_n(z,t)= e^{-iEt/\hbar}\,
\varphi^{\pm}_{E,n}(z)
\end{equation}
with
\begin{equation}
\varphi^{\pm}_{E,n}(z) = A^{\pm}_n\,
\mathrm{Ai}(\tilde{z}-\tilde{h}_E+\tilde{h}^{\pm}_n) + B^{\pm}_n\,
\mathrm{Bi}(\tilde{z}-\tilde{h}_E+\tilde{h}^{\pm}_n),
\end{equation}
where $A^{\pm}_n$ et $B^{\pm}_n$ are complex coefficients and
$h^{\pm}_n= V^\pm_n/m\mathcal{G}$, \emph{i.e.}
\begin{subequations}
\begin{align}
    h^{+}_n & = h_{\mathrm{int}} \tan \theta_n, \\
    h^{-}_n & = -h_{\mathrm{int}} \cot \theta_n,
\end{align}
\end{subequations}

From Eq.~(\ref{basis}), we may express the wave function components
of the atom-field state inside the cavity over the non-coupled state
basis. We have
\begin{subequations}
\begin{align}
    \psi^a_n(z,t) & = \cos \theta_n \psi^+_n(z,t) - \sin \theta_n
    \psi^-_n(z,t), \\
    \psi^b_{n+1}(z,t) & = \sin \theta_n \psi^+_n(z,t) + \cos \theta_n
    \psi^-_n(z,t).
\end{align}
\end{subequations}

This allows us to find the wave function components of the
eigenstate $|\phi_E\rangle$ over the entire $z$ axis. The relations
(\ref{psiapsibn}) hold with
\begin{align}\label{glob3}
&\varphi^a_{E,n}(z)= \left\{\begin{array}{ll}
a_{E,n}^a\:\mathrm{Ai}(\tilde{z}-\tilde{h}_E) &\hspace{0.4cm}z>L,\vspace{5pt}\\
\varphi^a_{E,n}(z)\big|_\mathrm{c}&\hspace{0.4cm}0\leqslant  z\leqslant  L,\vspace{5pt}\\
\mathrm{U}(\tilde{z}-\tilde{h}_E)+d_{E,n}^a\:\mathrm{D}(\tilde{z}-\tilde{h}_E)&\hspace{0.4cm}z<0,
\end{array}\right.\\
&\varphi^b_{E,n+1}(z)= \left\{\begin{array}{ll}\label{glob4}
a_{E,n+1}^b\:\mathrm{Ai}(\tilde{z}-\tilde{h}_E+\tilde{\delta}) &\hspace{0.4cm}z>L,\vspace{5pt}\\
\varphi^b_{E,n+1}(z)\big|_\mathrm{c}&\hspace{0.4cm}0\leqslant {z}\leqslant {L},\vspace{5pt}\\
d_{E,n+1}^b\:\mathrm{D}(\tilde{z}-\tilde{h}_E+\tilde{\delta})&\hspace{0.4cm}z<0,
\end{array}\right.
\end{align}
and
\begin{subequations}
\label{phianzphibnz}
\begin{align}
& \varphi^a_{E,n}(z)\big|_\mathrm{c} =\nonumber\\ &
\cos\theta_{n}\Big(A^{+}_n\,
\mathrm{Ai}(\tilde{z}-\tilde{h}_E+\tilde{h}^{+}_{n}) + B^{+}_n\,
\mathrm{Bi}(\tilde{z}-\tilde{h}_E+\tilde{h}^{+}_{n})\Big) \nonumber
\\ & -\sin\theta_{n}\Big(A^{-}_n\,
\mathrm{Ai}(\tilde{z}-\tilde{h}_E+\tilde{h}^{-}_{n}) + B^{-}_n\,
\mathrm{Bi}(\tilde{z}-\tilde{h}_E+\tilde{h}^{-}_{n})\Big),
\\
& \varphi^b_{E,n+1}(z)\big|_\mathrm{c} =\nonumber\\ &
\sin\theta_{n}\Big(A^{+}_n\,
\mathrm{Ai}(\tilde{z}-\tilde{h}_E+\tilde{h}^{+}_{n}) + B^{+}_n\,
\mathrm{Bi}(\tilde{z}-\tilde{h}_E+\tilde{h}^{+}_{n})\Big) \nonumber
\\ &
 +\cos\theta_{n}\Big(A^{-}_n\,
\mathrm{Ai}(\tilde{z}-\tilde{h}_E+\tilde{h}^{-}_{n}) + B^{-}_n\,
\mathrm{Bi}(\tilde{z}-\tilde{h}_E+\tilde{h}^{-}_{n})\Big).
\end{align}
\end{subequations}

The coefficients $a_{E,n}^a$, $a_{E,n+1}^b$, $d_{E,n}^a$,
$d_{E,n+1}^b$, $A^+_n$, $A^-_n$, $B^+_n$, and $B^-_n$ in
expressions (\ref{glob3})--(\ref{phianzphibnz}) are found by
imposing the continuity conditions on the wave function and its
first derivative at the cavity interfaces.

\section{Induced emission probability}
\label{PemSection}

In this section, we derive a formula for the induced emission
probability of a photon inside the cavity. To this end, we will use
simple physical arguments rather than a rigorous treatment based on
wavepackets like in Ref.~\cite{Bas05a} as these two approaches give
the same result. According to Eqs.~(\ref{glob3})--(\ref{glob4}) a
monoenergetic excited atom coming upwards with unit amplitude upon
the cavity that contains $n$ photons is found to move back downwards
below the cavity in the $|a\rangle$ state with amplitude
${d}_{E,n}^a$ and similarly in the $|b\rangle$ state with amplitude
${d}_{E,n+1}^b$. As the probability current density of the downward
wave does not depend on the energy (see Eq.~(\ref{current})), the
induced emission probability $\mathcal{P}_{\mathrm{em}}(n)$ of a
photon inside the cavity containing initially $n$ photons is simply
given by
\begin{equation}\label{Pem}
    \mathcal{P}_{\mathrm{em}}(n)=\left|{d}_{E,n+1}^b\right|^2,
\end{equation}
and one has
\begin{equation}
    \left|{d}_{E,n+1}^b\right|^2+\left|{d}_{E,n}^a\right|^2=1.
\end{equation}

As shown in Appendix A, this probability can be expressed as a
function of the sole wave function component $\varphi^{a}_{E,n}$ of
the excited atom inside the cavity through the relation
\begin{equation}\label{pemphia}
\mathcal{P}_{\mathrm{em}}(n)=\pi^2\tilde{h}_{\mathrm{int}}^2
\left|\int_{-\infty}^{+\infty}
u(\tilde{z})\,\varphi^{a}_{E,n}(\tilde{z})\,
\mathrm{Ai}\big(\tilde{z}-\tilde{h}_E+\tilde{\delta}\big)~\mathrm{d}\tilde{z}\right|^2,
\end{equation}
valid for any mode function $u(z)$. This relation will help us
giving a physical interpretation to our results.

Another relation involving the induced emission probability follows
from a certain symmetry of Hamiltonian (\ref{Hamiltonian}). Denoting
the induced emission probability for a total energy $E$ and a
detuning $\delta$ by
$\mathcal{P}_{\mathrm{em}}(\tilde{h}_E,\tilde{\delta})$, it can be
easily proven that for any mode function $u(z)$ we have
\begin{equation}\label{relpem}
    \mathcal{P}_{\mathrm{em}}(\tilde{h}_E,-\tilde{\delta})=
\mathcal{P}_{\mathrm{em}}(\tilde{h}_E+\tilde{\delta},\tilde{\delta}).
\end{equation}

We study hereafter the properties of the induced emission
probability $\mathcal{P}_{\mathrm{em}}(n)$ depending on whether the
atomic kinetic energy (or the lack of this energy) at the cavity
interfaces is much higher than the interaction energy ($|h_E| \gg
h_{\mathrm{int}}$ and $|h_E - L| \gg h_{\mathrm{int}}$) or much
lower. We call each case the classical and quantum regime,
respectively.\\

\subsection{Classical regime~: Rabi limit}

At resonance ($\omega=\omega_0$), it has been shown in
Ref.~\cite{Bas05a} that the induced emission probability (\ref{Pem})
reduces, in the classical regime ($|h_E| \gg h_{\mathrm{int}}$ and
$|h_E - L| \gg h_{\mathrm{int}}$), to the well known Rabi
formula~\cite{Sch01}
\begin{equation} \label{classres2}
\mathcal{P}_{\mathrm{em}}(n) = \sin^2\left(\frac{\Omega_n
\tau}{2}\right),
\end{equation}
where $\tau$ is the classical transit time of an atom of energy $E$
through the cavity, indicating that way that the quantization of the
atomic motion is unnecessary in this particular regime. The formula
generalizing Eq.~(\ref{classres2}) in the presence of a detuning
$\delta$ can be easily derived using the classical results of the
interaction of a two-level atom with a quantized electromagnetic
field~\cite{Sch01} and is given by
\begin{align}\label{clasres}
& \mathcal{P}_{\mathrm{em}}(n) =
\left(\frac{\Omega_n}{\Omega'_n}\right)^2\,\times\nonumber\\ &
\left\{\begin{array}{ll}
     4\sin^2\frac{\Omega'_n \tau_2}{2}\,\times\\
   \left[\cos\frac{\Omega'_n \tau_2}{2}\cos\frac{\delta T}{2}-
   \frac{\delta}{\Omega'_n}\sin\frac{\Omega'_n \tau_2}{2}\sin\frac{\delta
   T}{2}\right]^2 & \hspace{0.5cm}L<h_E,\vspace{5pt}\\
     \sin^2 \frac{\Omega'_n  \tau_1}{2} & \hspace{0.5cm}0\leqslant {h}_E\leqslant {L},\vspace{5pt}\\
     0 & \hspace{0.5cm}h_E<0,
     \end{array}\right.
\end{align}
where $\tau_1$ and $\tau_2$ are the classical interaction times for
$0\leqslant {h}_E\leqslant {L}$ and $L<h_E$, respectively. For
$\tilde{h}_E<0$, the atom does not reach the cavity and no induced
emission process can occur (the classical interaction time $\tau$ is
null). For $0<\tilde{h}_E<\tilde{L}$, the atom turns back inside the
cavity and the interaction time $\tau_1$ does not depend on the
cavity length $L$. For $\tilde{h}_E>\tilde{L}$, the atom passes
twice inside the cavity (upwards and downwards) and the interaction
time $\tau_2$ is $L$ dependent. We will denote the time which passed
between the first and the second interaction by $T$. In terms of
dimensionless variables, these various times are given by
\begin{align}
    & \Omega'_n\tau_2=\sqrt{\tilde{\delta}^2+
    4\:\tilde{h}_{\mathrm{int}}^2}\left(\sqrt{\tilde{h}_E}-\sqrt{\tilde{h}_E-\tilde{L}}\:\right),\\
  &   \Omega'_n\tau_1=2\sqrt{\tilde{\delta}^2+
    4\:\tilde{h}_{\mathrm{int}}^2}\,\sqrt{\tilde{h}_E},\\
   &  \delta T=2\tilde{\delta}\sqrt{\tilde{h}_E-\tilde{L}}.
\end{align}

\begin{figure*}
\centering
\includegraphics*[width=13cm]{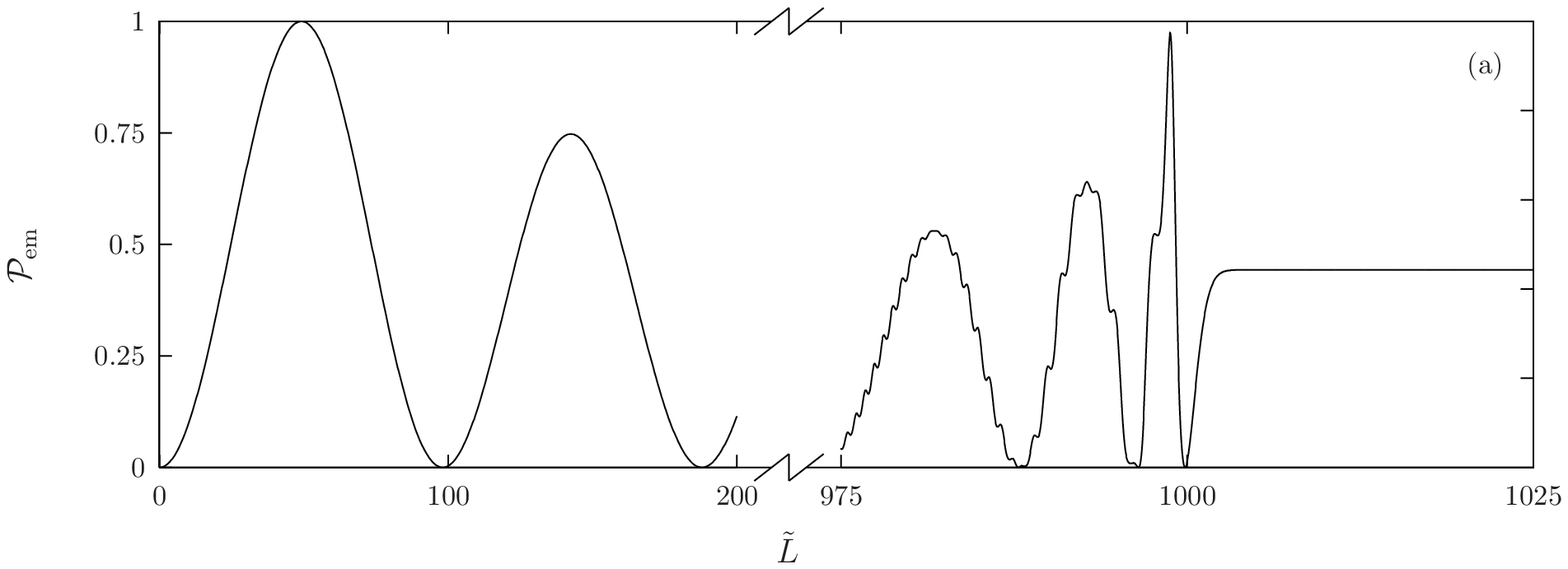}\vspace{10pt}
\includegraphics*[width=13cm]{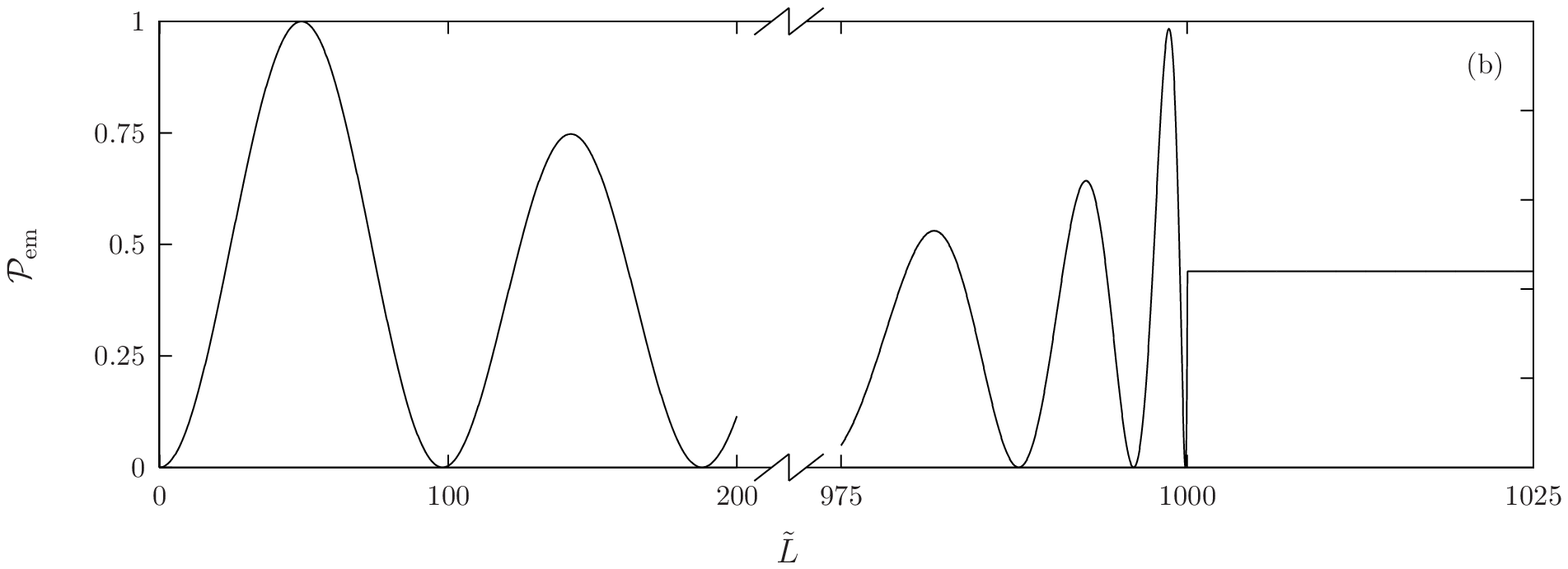}
\caption{Induced emission probability $\mathcal{P}_{\mathrm{em}}$
with respect to the cavity length (a) with and (b) without
quantization of the atomic motion (classical regime:
$\tilde{h}_E=1000$, $\tilde{h}_{\mathrm{int}}=1$ and
$\tilde{\delta}=-0.2$).}\label{pemahElong}
\end{figure*}

We illustrate in Figs.~\ref{pemahElong}(a) and \ref{pemahElong}(b)
the induced emission probability $\mathcal{P}_{\mathrm{em}}\equiv
\mathcal{P}_{\mathrm{em}}(n)$ with respect to the cavity length for
$\tilde{h}_E \gg \tilde{h}_{\mathrm{int}}$, respectively with and
without quantization of the atomic motion (Eqs.~(\ref{Pem}) and
(\ref{clasres}) respectively). We observe that both results are in
good agreement, except in the region where $\tilde{L} \approx
\tilde{h}_E$. In this region, $|\tilde{h}_E - \tilde{L}|$ is not
much greater than $\tilde{h}_{\mathrm{int}}$ and we are out of the
validity domain of the approximated expression (\ref{clasres}). The
small additional oscillations observed in the induced emission
probability curve where $\tilde{L} \approx \tilde{h}_E$ are pure
quantum effects originating from the quantization of the atomic
motion. These effects arise when the cavity length fits nearly the
classical turning points of the atoms in the gravitational field. In
this case, the potential barrier and well $V_n^{\pm}(z)$ play a
significant role as the kinetic energy of the atoms is very small
exactly where the potential energy exhibits strong variations.

\subsection{Quantum regime~: mazer limit}

\begin{figure}
\begin{center}
\includegraphics[width=.95\linewidth]{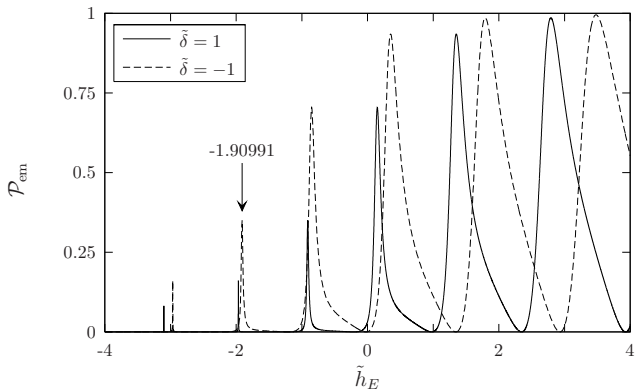}
\end{center}
\caption{Induced emission probability with respect to the total
atomic energy for $\tilde{L}=10$, $\tilde{h}_{\mathrm{int}}=10$ and
$n=0$.}\label{pemahE}
\end{figure}

\begin{figure}
\begin{center}
\includegraphics[width=.95\linewidth]{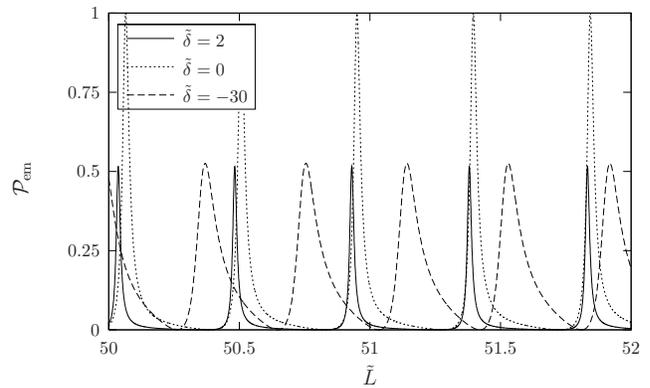}
\end{center}
\caption{Induced emission probability with respect to the cavity
length for $\tilde{h}_E=1$, $\tilde{h}_{\mathrm{int}}=100$ and
$n=0$.}\label{pemaL}
\end{figure}

In the quantum regime, the induced emission probability
$\mathcal{P}_{\mathrm{em}}$ exhibits a completely different
behavior. This is first illustrated in Figs.~\ref{pemahE} and
\ref{pemaL} that show $\mathcal{P}_{\mathrm{em}}$ as a function of
the total atomic energy and the cavity length, respectively, for
nonvanishing detunings. Contrary to the classical regime, narrow
resonances are presently observed, even in the negative energy
domain [$E<0$, \emph{i.e.}, $h_E<0$, see Eq.~(\ref{he})] where
classically the atom does not reach the cavity and cannot therefore
emit any photon in there. The existence of negative energy
resonances reflects the fact that the atom can still interact with
the cavity field for $h_E<0$ because of the \emph{tunnel effect}.
Indeed, we have shown in~\cite{Bas05a} that at resonance
($\delta=0$) the total potential felt by the atom in interaction
with the cavity possesses quasibound states of negative energy.
Whenever such a state energy matches the atomic $E$ value, the atom
can enter the cavity by tunnel effect and emit a photon in there. In
the presence of a detuning, the emission probability resonances
still exist but with a modified position, amplitude, and width.

\begin{figure}
\centering \subfigure[$\tilde{h}_{E}=-1.90991$
$[\mathcal{P}_{\mathrm{em}}\simeq
0.35{]}$]{\includegraphics*[width=7.5cm]{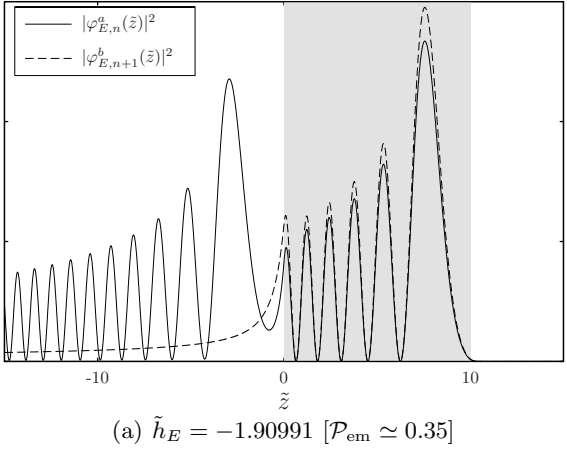}\label{densprobabmesaa}}
\subfigure[ $\tilde{h}_{E}=-2.2$ $[\mathcal{P}_{\mathrm{em}}\simeq
0{]}$]{\includegraphics*[width=7.5cm]{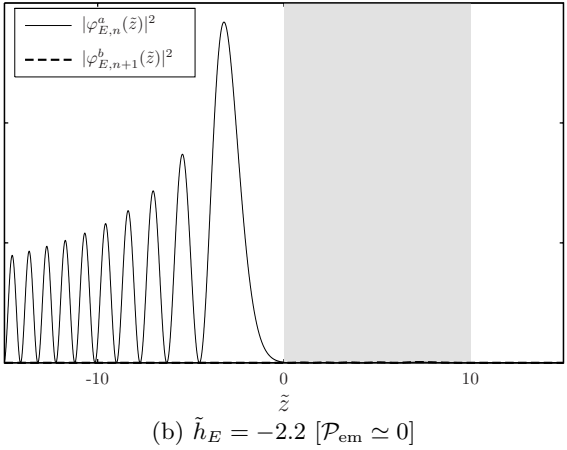}\label{densprobabmesab}}
 \caption[]{Atomic probability densities
$|\varphi^a_{E,n}(\tilde{z})|^2$ and
$|\varphi^b_{E,n+1}(\tilde{z})|^2$ for $\tilde{L}=10$,
$\tilde{h}_{\mathrm{int}}=10$, $\tilde{\delta}=-1$,
 and $n=0$. The grey area shows the cavity region.}\label{densprobabmesa}
\end{figure}

From a physical point of view, the behavior of the stationary wave
functions $\varphi^a_{E,n}(\tilde{z})$ and
$\varphi^b_{E,n+1}(\tilde{z})$ [Eqs.~(\ref{glob3}) and
(\ref{glob4})] differs strongly depending on whether the value of
the total atomic energy $E$ gives rise or not to a resonance in the
emission probability curve [see Fig.~\ref{pemahE}]. This is
illustrated in Figs.~\ref{densprobabmesaa} and \ref{densprobabmesab}
that show the atomic probability densities
$|\varphi^a_{E,n}(\tilde{z})|^2$ and
$|\varphi^b_{E,n+1}(\tilde{z})|^2$ inside and outside the cavity
(grey and white area respectively). In the first case, the atom is
launched upwards to the cavity with a total energy
$\tilde{h}_{E}=-1.90991$ matching a resonance in the emission
probability curve [see Fig.~\ref{pemahE}] and the stationary wave
functions $\varphi^a_{E,n}(\tilde{z})$ and
$\varphi^b_{E,n+1}(\tilde{z})$ take a significant value inside the
cavity. Consequently, a strong interaction between the atom and the
cavity field can occur and the induced emission probability is
significant, according to Eq.~(\ref{pemphia}). When considering a
realistic atomic wave packet, this effect is of course restricted to
the components of the wave packet for which $\tilde{h}_E$ falls
within the width of the resonance. Inversely, if the total atomic
energy does not match a resonance energy
(Fig.~\ref{densprobabmesab}), the stationary wave functions are
almost zero inside the cavity and no atom-field interaction can take
place. In this case, the induced emission probability drops down to
zero according to Eq.~(\ref{pemphia}).

\subsubsection{Peak position}
\begin{figure}
\begin{center}
\includegraphics[width=.95\linewidth]{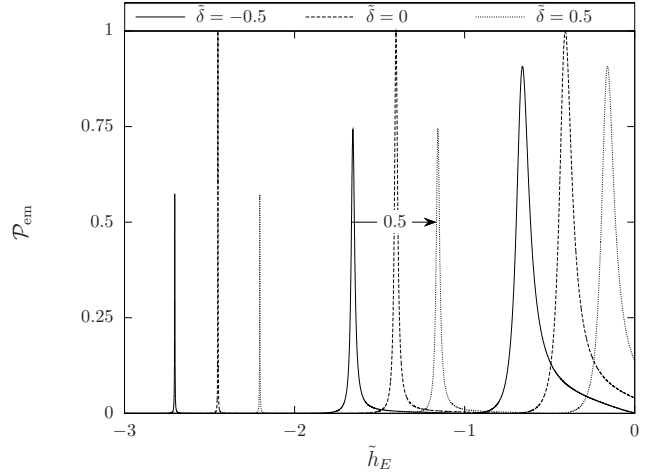}
\end{center}
\caption{Induced emission probability with respect to $\tilde{h}_E$
for $\tilde{h}_{\mathrm{int}}=10$, $n=0$, and
$\tilde{L}\gg\tilde{h}_E+\tilde{h}_{\mathrm{int}}$.}\label{pemahEpos}
\end{figure}

Figure~\ref{pemahEpos} shows the induced emission probability for a
positive, a negative and a null detuning. We observe that for
positive detunings, the emission probability peaks move towards
increasing energies while for negative ones it is the opposite.
According to Eq.~(\ref{relpem}), peaks for opposite detunings
$\pm\tilde{\delta}$ are separated in energy by $\tilde{\delta}$
($0.5$ in the case of Fig.~\ref{pemahEpos}). From
Fig.~\ref{pemahEpos} and further calculations, it turns out that the
peak energy shift caused by a detuning $\pm\delta$ is in very good
approximation given by $\pm\tilde{\delta}/2$.

\subsubsection{Peak amplitude and width}

\begin{figure}
\begin{center}
\includegraphics[width=.95\linewidth]{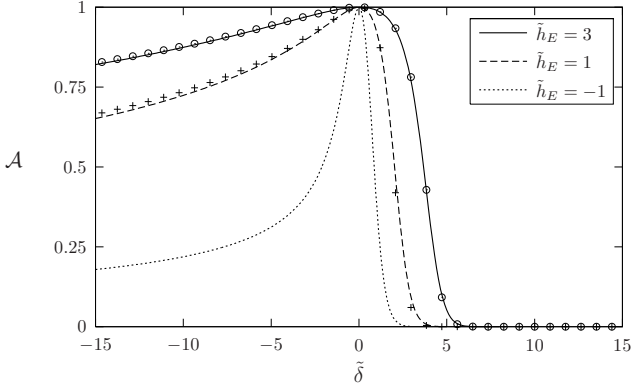}
\end{center}
\caption{Amplitude $\mathcal{A}$ of the resonances in the induced
emission probability with respect to $\tilde{\delta}$ for
$\tilde{h}_{\mathrm{int}}=100$ and $n=0$. Circles and crosses are
the values computed from (\ref{amplianalytic}) for $\tilde{h}_E=3$
and $\tilde{h}_E=1$ respectively.}\label{ampahE}
\end{figure}

Figure~\ref{ampahE} shows how the peak amplitude evolves with
respect to the detuning for various fixed values of the total atomic
energy. In contrast to the classical regime [see
Eq.~(\ref{clasres})], the curves present a strong asymmetry with
respect to the sign of the detuning. This results from the atomic
energy variation $\hbar \delta$ when the atom emits a photon inside
the cavity (see Sec.\ \ref{wfunctions}). The smaller the total
atomic energy is, the more pronounced this asymmetry is. The induced
emission probability tends very rapidly to zero for positive
detunings, in contrast to what happens for negative detunings. This
behavior is already present in Eq.~(\ref{pemphia}) where the Airy Ai
function decreases much faster for large positive arguments than for
large negative arguments as can be seen from the asymptotic
expansions~\cite{Abr70}
\begin{align}
& \mathrm{Ai}(-z) =
    \frac{z^{-1/4}}{\sqrt{\pi}} \sin\left(\xi+\frac{\pi}{4}\right)
    +\mathcal{O}\big(z^{-7/4}\big)\\
& \mathrm{Ai}(z) =
    \frac{1}{2\sqrt{\pi}}\:z^{-1/4}e^{-\xi}+\mathcal{O}\big(z^{-7/4}e^{-\xi}\big)
\end{align}
valid for $|z|\gg 1$ and where $\xi=\frac{2}{3}z^{3/2}$.

This asymmetry was also present in the absence of
gravity~\cite{Bas03b} [horizontal geometry, see Fig.~\ref{hm}]. In
Ref.~\cite{Bas03b}, it had been shown that the induced emission
probability is strictly null for $\hbar\delta>E$. This results
merely from energy conservation. If the initial atomic kinetic
energy $E$ is lower than $\hbar \delta$, the transition $|a,n\rangle
\rightarrow |b,n+1\rangle$ cannot take place (as it would remove
$\hbar \delta$ from the kinetic energy) and no photon can be emitted
inside the cavity. In this case the emission process is completely
blocked. In the presence of gravity, there doesn't exist such a
threshold. For instance, we can see in Fig.~\ref{ampahE} that the
peak amplitude for $\tilde{h}_E=3$ takes significant values beyond
$\tilde{h}_{\delta}=3$. This is not really surprising since quantum
mechanically a particle of energy $E$ moving freely in the gravity
field is described by the wave function $\varphi(\tilde{z})=N
\mathrm{Ai}(\tilde{z}-\tilde{h}_E)$ and has a nonvanishing
probability of being found in the nonclassical region $z>h_E$. This
probability decreases however very fast as a function of the height
$z$. This enables us to understand qualitatively why the peak
amplitude decreases very rapidly but not sharply (no threshold as in
the case of the horizontal mazer) as a function of the detuning for
$\tilde{\delta}>\tilde{h}_E$.

\begin{figure}
\begin{center}
\includegraphics[width=.95\linewidth]{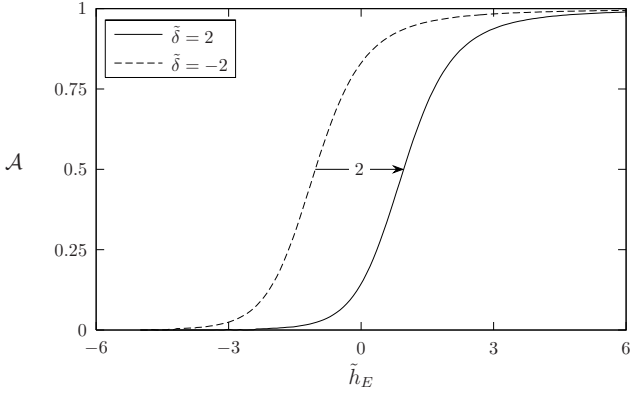}
\end{center}
\caption{Amplitude $\mathcal{A}$ of the resonances in the induced
emission probability with respect to $\tilde{h}_E$ for
$\tilde{h}_{\mathrm{int}}=100$ and $n=0$.}\label{ampahd}
\end{figure}

Figure~\ref{ampahd} shows the peak amplitude with respect to the
total atomic energy for two opposite values of the detuning
$\tilde{\delta}=\pm 2$. According to Eq.~(\ref{relpem}), the curves
have the same shape and are merely shifted in energy by
$\tilde{h}_{E}=2$.

We have observed that the peak amplitude only slightly depends on
the interaction height $\tilde{h}_{\mathrm{int}}$ and that it is
well approximated by the following simple analytical formula
\begin{equation}\label{amplianalytic}
    \mathcal{A}=
    \frac{4}{\pi^2}\Big|\mathrm{U}(-\tilde{h}_E)\mathrm{D}'(\tilde{\delta}-\tilde{h}_E)
    -\mathrm{D}(\tilde{\delta}-\tilde{h}_E)
    \mathrm{U}'(-\tilde{h}_E)\Big|^{-2}
\end{equation}

In Fig.~\ref{ampahE}, we compare the peak amplitude derived from
Eq.~(\ref{Pem}) and computed from Eq.~(\ref{amplianalytic}),
respectively. For $\tilde{h}_E=3$ and higher values of the energy, a
good agreement is observed. However, this agreement lowers as the
energy decreases.

Concerning the peak width, we have observed that negative detunings
increase the peak width while positive ones decrease it (see for
example Fig.~\ref{pemaL}).

\subsubsection{Cavity length effects}

\begin{figure}
\begin{center}
\includegraphics[width=.95\linewidth]{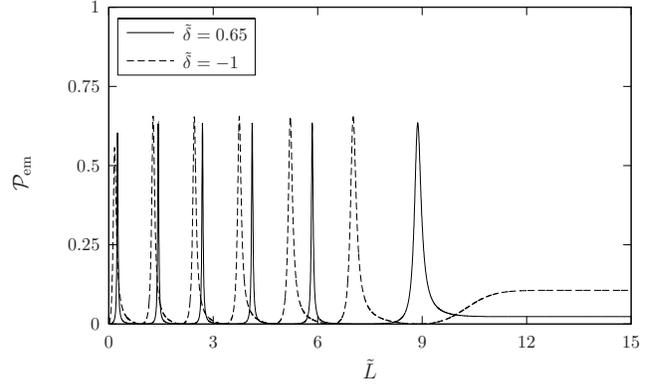}
\end{center}
\caption{Induced emission probability with respect to the cavity
length for $\tilde{h}_E=-1$, $\tilde{h}_{\mathrm{int}}=10$ and
$n=0$.}\label{pemaL2}
\end{figure}

In the classical regime and in the positive energy domain, the
induced emission probability does not depend on the cavity length
$L$ as soon as this length is greater than $h_E$ (see
Eq.~(\ref{clasres})). In the quantum regime, this statement must be
revised and holds only when $L \gtrsim h_E + h^+_n$. This is
illustrated in Fig.~\ref{pemaL2} which shows the induced emission
probability $\mathcal{P}_{\mathrm{em}}$ with respect to the cavity
length for $\tilde{h}_E = -1$ and $\tilde{h}_{\mathrm{int}} = 10$.
For $\tilde{h}_{\delta} = -1$, $\mathcal{P}_{\mathrm{em}}$ starts to
be constant only beyond $\tilde{L} = \tilde{h}_E + \tilde{h}^+_n
\simeq 8.512$.

\subsubsection{Detuning sensitiveness}

\begin{figure}
\begin{center}
\includegraphics[width=.95\linewidth]{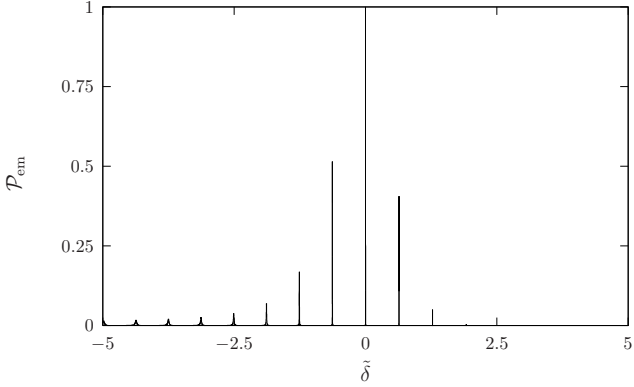}
\end{center}
\caption{Induced emission probability with respect to the detuning
for $\tilde{h}_E=-2.5$, $\tilde{h}_{\mathrm{int}}=100$, $n=0$, and
$\tilde{L}\gg\tilde{h}_E+\tilde{h}_{\mathrm{int}}$. For rubidium
atoms, the central peak has a width of only 0.3~Hz.}\label{pemahd}
\end{figure}

Figure~\ref{pemahd} displays the induced emission probability with
respect to the detuning. For realistic experimental
parameters~\cite{Lof97} and rubidium atoms, these resonances may
even become extremely narrow. Their width amounts only $0.3$~Hz for
$\tilde{h}_E=-2.5$ and $\tilde{h}_{\mathrm{int}}=100$. This could
define very useful metrology devices (atomic clocks for example)
based on a single cavity and with better performances than what is
usually obtained in the well known Ramsey configuration
\cite{Cla91}.

\subsection{Sine and Gaussian mode functions}
\begin{figure}
\centering
\includegraphics*[width=.95\linewidth]{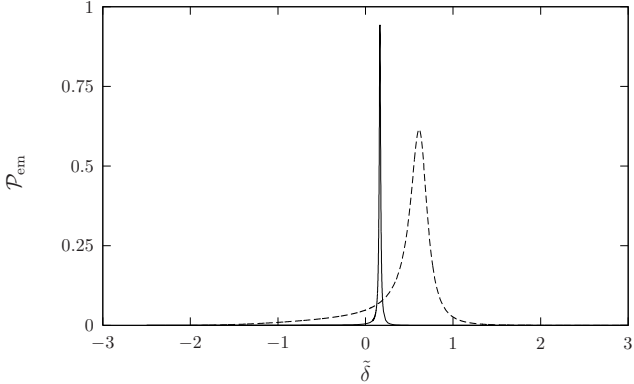}
\caption[]{Induced emission probability with respect to the detuning
for a sine mode with $\tilde{h}_{\mathrm{int}}=100$, $\tilde{L}=10$
(solid line) and
 for a gaussian mode
with $\tilde{h}_{\mathrm{int}}=10$, $\tilde{L}=1$ (dashed line)
($\tilde{h}_E=-2$ and $n=0$ in both cases).}\label{pemsingauss}
\end{figure}

The narrow resonances of the induced emission probability observed
in the quantum regime are not restricted to the case of the mesa
mode investigated so far here. Using numerical procedures (see
Appendix B), we have computed the solutions
$\varphi_{E,n}^{a}(\tilde{z})$ and $\varphi_{E,n+1}^{b}(\tilde{z})$
of the stationary Schr\"odinger equation (\ref{sysreduitmv}) for
sine and gaussian mode functions [$u(z) = \sin(\pi z / L)$ for $0 <
z < L$, 0 elsewhere and $u(z) = e^{-(z-L/2)^2/2\sigma^2}$ with
$\sigma = L/\sqrt{2}$, respectively]. We show in
Fig.~\ref{pemsingauss} the related induced emission probability
$\mathcal{P}_{\mathrm{em}}$ with respect to the detuning. Again, we
observe sharp resonances whose width decreases with increasing
interaction height ($h_{\mathrm{int}}$).

\begin{figure}
\begin{center}
\includegraphics[width=.95\linewidth]{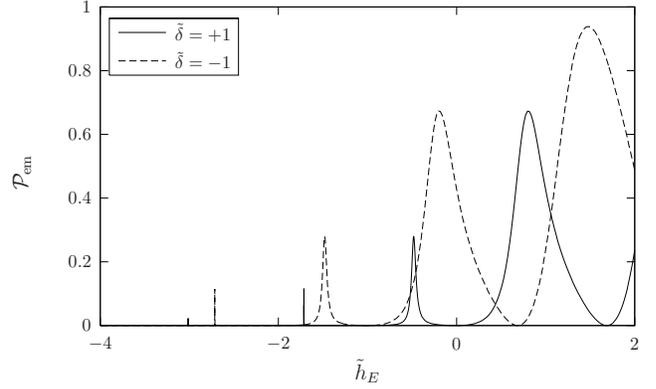}
\end{center}
\caption{Induced emission probability with respect to the total
atomic energy for a sine mode and $\tilde{h}_{\mathrm{int}}=10$,
$\tilde{L}=10$ and $n=0$.}\label{pemsina}
\end{figure}

As any mode function will give rise to such resonances (provided
$\tilde{h}_{\textrm{int}}$ is large enough), the resonances in the
induced emission probability will be a very general feature of the
vertical micromaser in the cold atom regime, in contrast to what
happens with the horizontal micromaser where the induced emission
probability characteristics are much more mode
dependent~\cite{Lof97}. This defines a very interesting property as
it should help the experimenters to observe the particular behavior
of the induced emission probability in the quantum regime, whatever
the exact mode function of the cavity. Another nice illustration of
the symmetry relation (\ref{relpem}) is displayed in
Fig.~\ref{pemsina}, this time for a sine mode function.

\section{Summary}
\label{SummarySection}
 In this paper we have presented the quantum
theory of the vertical mazer for a non resonant atom-field
interaction. We have obtained analytical expressions for the wave
function in the special case of a constant cavity mode function. The
properties of the induced emission probability in the presence of a
detuning have been discussed both in the classical and the quantum
regimes. The classical results are well recovered for hot atoms for
which the quantization of the atomic motion is unnecessary. We have
shown that atoms which classically would not reach the interaction
region are able to emit a photon inside the cavity. In the quantum
regime, the mazer properties are not symmetric with respect to the
sign of the detuning and the system exhibits a sharp response as a
function of the detuning. The induced emission probability has also
been computed for sine and gaussian mode functions (see Appendix B
for the numerical procedure we have developed).

\section{Acknowledgments}
\label{Acknowledgments} This work has been supported by the Belgian
Institut Interuniversitaire des Sciences Nucl\'eaires (IISN). J.M.\
thanks the Belgian FRIA for financial support and the University of
Li\`{e}ge where part of this work has been done. J.M.\ also thanks
the French ANR (project INFOSYSQQ, contract number ANR-05-JCJC-0072)
for funding.

\section*{Appendix A}
\setcounter{section}{1}
\renewcommand{\thesection}{\Alph{section}}
\renewcommand{\theequation}{A.\arabic{equation}}
\setcounter{equation}{0}

In this appendix, we derive Eq.~(\ref{pemphia}) which expresses the
induced emission probability as a function of the stationary wave
function of the excited atom inside the cavity. To this end, let us
recall the equations of motion (\ref{sysreduitmv}) for the wave
function components $\varphi^a_{E,n}$ and $\varphi^b_{E,n+1}$ (see
Eq.~(\ref{psiapsibn}))
\begin{subequations}\label{sys}
\begin{align}
        & \left(\frac{\mathrm{d}^2}{\mathrm{d}
        \tilde{z}^2}- \tilde{z}+ \tilde{h}_E\right)\varphi^{a}_{E,n}(\tilde{z}) = \tilde{h}_{\mathrm{int}}
        \,u(\tilde{z})\,\varphi^{b}_{E,n+1}(\tilde{z}),\label{sysa}\\
        & \left(\frac{\mathrm{d}^2}{\mathrm{d}
        \tilde{z}^2}-\tilde{z} + \tilde{h}_E - \tilde{\delta}\right)\varphi^{b}_{E,n+1}(\tilde{z})
        = \tilde{h}_{\mathrm{int}}
        \,u(\tilde{z})\,\varphi^{a}_{E,n}(\tilde{z}).\label{sysb}
\end{align}
\end{subequations}

To solve the problem numerically, the domain is restricted to
$[\tilde{z}_{\mathrm{min}},\tilde{z}_{\mathrm{max}}]$. Outside this
domain, the mode function $u(\tilde{z})$ is regarded as null.

The outgoing Green function of the stationary Schr\"odinger equation
of a particle of energy $E$ in the gravitational field, satisfying
\begin{equation}
    \left(\frac{\mathrm{d}^2}{\mathrm{d}
        \tilde{z}^2}- \tilde{z}+ \tilde{h}_E\right)G_{E}(\tilde{z}|\tilde{z}')=\delta(\tilde{z}-\tilde{z}'),
\end{equation}
is given by~\cite{Bra97b}
\begin{equation}\label{fctgreenlin}
G_E(\tilde{z}|\tilde{z}')= -i\pi
\,\mathrm{D}(\alpha_{E,-})\,\mathrm{Ai}(\alpha_{E,+})
\end{equation}
with
\begin{equation}
\alpha_{E,\pm}=-\tilde{h}_E+\frac{(\tilde{z}+\tilde{z}')\pm|\tilde{z}-\tilde{z}'|}{2}.
\end{equation}

On the one hand, considering the right-hand side term of
Eq.~(\ref{sysb}) as a source term and using Green function
(\ref{fctgreenlin}), the wave function component
$\varphi^{b}_{E,n+1}(\tilde{z})$ is given at
$\tilde{z}=\tilde{z}_{\mathrm{min}}$ by
\begin{multline}\label{as}
\varphi^b_{E,n+1}(\tilde{z}_{\mathrm{min}}) = -i\pi
\tilde{h}_{\mathrm{int}}\,\mathrm{D}\big(\tilde{z}_{\mathrm{min}}-\tilde{h}_E+\tilde{\delta}\big)\times\\
\int_{-\infty}^{+\infty}
u(\tilde{z})\,\varphi^{a}_{E,n}(\tilde{z})\,
\mathrm{Ai}\big(\tilde{z}-\tilde{h}_E+\tilde{\delta}\big)~\mathrm{d}\tilde{z}
\end{multline}
since by definition $u(\tilde{z})=0$ for
$\tilde{z}<\tilde{z}_{\mathrm{min}}$. On the other hand,
$\varphi^{b}_{E,n+1}(\tilde{z})$, which is continuous at
$\tilde{z}=\tilde{z}_{\mathrm{min}}$, is also given by
Eq.~(\ref{solb}). Identification with Eq.~(\ref{as}) then yields
\begin{equation}\label{dcoeffpem}
d_{E,n+1}^b=-i\pi \tilde{h}_{\mathrm{int}}
\,\int_{-\infty}^{+\infty}
u(\tilde{z})\,\varphi^{a}_{E,n}(\tilde{z})\,
\mathrm{Ai}\big(\tilde{z}-\tilde{h}_E+\tilde{\delta}\big)~\mathrm{d}\tilde{z}
\end{equation}
whose absolute value squared gives Eq.~(\ref{pemphia}).

\section*{Appendix B}
\setcounter{section}{2}
\renewcommand{\thesection}{\Alph{section}}
\renewcommand{\theequation}{B.\arabic{equation}}
\setcounter{equation}{0} \setlength{\arraycolsep}{2.5pt}

In this appendix, we describe the method we have developed to solve
numerically the system (\ref{sys}) of coupled differential equations
with imposed boundary conditions.

\subsection{Boundary conditions}
The continuity conditions of the wave function components
(\ref{sola}) and (\ref{solb}) and their first derivatives at the
cavity interfaces ($\tilde{z}_{\mathrm{min}}$ and $
\tilde{z}_{\mathrm{max}}$) read
\begin{equation}
\left\{\begin{array}{lll}
        \varphi^a_{E,n}(\tilde{z}_{\mathrm{min}})&=&\mathrm{U}\big(\tilde{z}_{\mathrm{min}}- \tilde{h}_E\big)
        +{d}^a_{E,n}\:\mathrm{D}\big(\tilde{z}_{\mathrm{min}}- \tilde{h}_E\big)\vspace{5pt}\\
        \frac{\mathrm{d}\varphi^a_{E,n}}{\mathrm{d}
        \tilde{z}}(\tilde{z}_{\mathrm{min}})&=&\mathrm{U}'\big(\tilde{z}_{\mathrm{min}}- \tilde{h}_E\big)
        +{d}^a_{E,n}\:\mathrm{D}'\big(\tilde{z}_{\mathrm{min}}- \tilde{h}_E\big)\vspace{5pt}\\
        \varphi^b_{E,n+1}(\tilde{z}_{\mathrm{min}})&=&{d}^b_{E,n+1}\:\mathrm{D}\big(\tilde{z}_{\mathrm{min}}+ \tilde{\delta}- \tilde{h}_E\big)\vspace{5pt}\\
        \frac{\mathrm{d}\varphi^b_{E,n+1}}{\mathrm{d}
        \tilde{z}}(\tilde{z}_{\mathrm{min}})&=&{d}^b_{E,n+1}\:\mathrm{D}'\big(\tilde{z}_{\mathrm{min}}+ \tilde{\delta}- \tilde{h}_E\big)
    \end{array}\right. \label{aeamin}
\end{equation}
and
\begin{equation}
    \left\{\begin{array}{lll}
        \varphi^a_{E,n}(\tilde{z}_{\mathrm{max}})&=&{a}^a_{E,n}\:\mathrm{Ai}\big(\tilde{z}_{\mathrm{max}}- \tilde{h}_E\big)\vspace{5pt}\\
        \frac{\mathrm{d}\varphi^a_{E,n}}{\mathrm{d}
        \tilde{z}}(\tilde{z}_{\mathrm{max}})&=&{a}^a_{E,n}\:\mathrm{Ai}'\big(\tilde{z}_{\mathrm{max}}- \tilde{h}_E\big)\vspace{5pt}\\
        \varphi^b_{E,n+1}(\tilde{z}_{\mathrm{max}})&=&{a}^b_{E,n+1}\:\mathrm{Ai}\big(\tilde{z}_{\mathrm{max}}+ \tilde{\delta}- \tilde{h}_E\big)\vspace{5pt}\\
        \frac{\mathrm{d}\varphi^b_{E,n+1}}{\mathrm{d}
        \tilde{z}}(\tilde{z}_{\mathrm{max}})&=&{a}^b_{E,n+1}\:\mathrm{Ai}'\big(\tilde{z}_{\mathrm{max}}+ \tilde{\delta}-
        \tilde{h}_E\big)
    \end{array}\right. \label{cimvd}
\end{equation}

\subsection{Resolution scheme}

We first introduce the vector
$\overrightarrow{\varphi}(\tilde{z})=\big({\varphi}^a_{E,n}(\tilde{z}),{\varphi}^b_{E,n+1}(\tilde{z})\big)^\mathrm{T}$.
Now, let $\overrightarrow{\varphi}_1(\tilde{z})$ and
$\overrightarrow{\varphi}_2(\tilde{z})$ be two vectors satisfying
(\ref{sysreduitmv}) with the boundary conditions (BC) (at
$\tilde{z}=\tilde{z}_{\mathrm{max}}$)
\begin{equation}
\setlength{\arraycolsep}{2.5pt}
  (1)  \left\{\begin{array}{lll}
    {\varphi}^a_{E,n}(\tilde{z}_{\mathrm{max}})&=&\mathrm{Ai}\big(\tilde{z}_{\mathrm{max}}- \tilde{h}_E\big)\vspace{5pt}\\
        \frac{\mathrm{d}{\varphi}^a_{E,n}}{\mathrm{d}\tilde{z}}(\tilde{z}_{\mathrm{max}})&=&\mathrm{Ai}'\big(\tilde{z}_{\mathrm{max}}- \tilde{h}_E\big)\vspace{5pt}\\
        {\varphi}^b_{E,n+1}(\tilde{z}_{\mathrm{max}})&=&0\vspace{5pt}\\
        \frac{\mathrm{d}{\varphi}^b_{E,n+1}}{\mathrm{d}\tilde{z}}(\tilde{z}_{\mathrm{max}})&=&0
    \end{array}\right.
\end{equation}
and
\begin{equation}
  (2)  \left\{\begin{array}{lll}
    {\varphi}^a_{E,n}(\tilde{z}_{\mathrm{max}})&=&0\vspace{5pt}\\
        \frac{\mathrm{d}{\varphi}^a_{E,n}}{\mathrm{d}\tilde{z}}(\tilde{z}_{\mathrm{max}})&=&0\vspace{5pt}\\
        {\varphi}^b_{E,n+1}(\tilde{z}_{\mathrm{max}})&=&\mathrm{Ai}\big(\tilde{z}_{\mathrm{max}}+ \tilde{\delta}- \tilde{h}_E\big)\vspace{5pt}\\
        \frac{\mathrm{d}{\varphi}^b_{E,n+1}}{\mathrm{d}\tilde{z}}(\tilde{z}_{\mathrm{max}})&=&\mathrm{Ai}'\big(\tilde{z}_{\mathrm{max}}+ \tilde{\delta}- \tilde{h}_E\big)
    \end{array}\right.
\end{equation}
The most general solution consistent with BC (\ref{cimvd}) is then
given by
\begin{equation}\label{phirightmv}
    \overrightarrow{\varphi}_{\mathrm{right}}(\tilde{z})={a}^a_{E,n}\,\overrightarrow{\varphi}_1(\tilde{z})+ {a}^b_{E,n+1}\,\overrightarrow{\varphi}_2(\tilde{z})
\end{equation}
where ${a}^a_{E,n}$ and ${a}^b_{E,n+1}$ are constants.

Next, let $\overrightarrow{\varphi}_3(\tilde{z})$,
$\overrightarrow{\varphi}_4(\tilde{z})$ and
$\overrightarrow{\varphi}_5(\tilde{z})$ be three vectors satisfying
(\ref{sysreduitmv}) with the boundary conditions (at
$\tilde{z}=\tilde{z}_{\mathrm{min}}$)
\begin{equation}
   (3) \left\{\begin{array}{lll}
    {\varphi}^a_{E,n}(\tilde{z}_{\mathrm{min}})&=&\mathrm{U}\big(\tilde{z}_{\mathrm{min}}- \tilde{h}_E\big)\vspace{5pt}\\
        \frac{\mathrm{d}{\varphi}^a_{E,n}}{\mathrm{d}\tilde{z}}(\tilde{z}_{\mathrm{min}})&=&\mathrm{U}'\big(\tilde{z}_{\mathrm{min}}- \tilde{h}_E\big)\vspace{5pt}\\
        {\varphi}^b_{E,n+1}(\tilde{z}_{\mathrm{min}})&=&0\vspace{5pt}\\
        \frac{\mathrm{d}{\varphi}^b_{E,n+1}}{\mathrm{d}\tilde{z}}(\tilde{z}_{\mathrm{min}})&=&0
    \end{array}\right. ,
\end{equation}
\begin{equation}
   (4) \left\{\begin{array}{lll}
    {\varphi}^a_{E,n}(\tilde{z}_{\mathrm{min}})&=&\mathrm{D}\big(\tilde{z}_{\mathrm{min}}- \tilde{h}_E\big)\vspace{5pt}\\
        \frac{\mathrm{d}{\varphi}^a_{E,n}}{\mathrm{d}\tilde{z}}(\tilde{z}_{\mathrm{min}})&=&\mathrm{D}'\big(\tilde{z}_{\mathrm{min}}- \tilde{h}_E\big)\vspace{5pt}\\
        {\varphi}^b_{E,n+1}(\tilde{z}_{\mathrm{min}})&=&0\vspace{5pt}\\
        \frac{\mathrm{d}{\varphi}^b_{E,n+1}}{\mathrm{d}\tilde{z}}(\tilde{z}_{\mathrm{min}})&=&0
    \end{array}\right.
\end{equation}
and
\begin{equation}
   (5) \left\{\begin{array}{lll}
    {\varphi}^a_{E,n}(\tilde{z}_{\mathrm{min}})&=&0\vspace{5pt}\\
        \frac{\mathrm{d}{\varphi}^a_{E,n}}{\mathrm{d}\tilde{z}}(\tilde{z}_{\mathrm{min}})&=&0\vspace{5pt}\\
        {\varphi}^b_{E,n+1}(\tilde{z}_{\mathrm{min}})&=&\mathrm{D}\big(\tilde{z}_{\mathrm{min}}+\tilde{\delta}- \tilde{h}_E\big)\vspace{5pt}\\
        \frac{\mathrm{d}{\varphi}^b_{E,n+1}}{\mathrm{d}\tilde{z}}(\tilde{z}_{\mathrm{min}})&=&\mathrm{D}'\big(\tilde{z}_{\mathrm{min}}+\tilde{\delta}- \tilde{h}_E\big)
    \end{array}\right.
\end{equation}

The most general solution consistent with BC (\ref{aeamin}) is then given by
\begin{equation}\label{phileftmv}
    \overrightarrow{\varphi}_{\mathrm{left}}(\tilde{z})=\overrightarrow{\varphi}_3(\tilde{z})+
    d^a_{E,n}\,\overrightarrow{\varphi}_4(\tilde{z})+d^b_{E,n+1}\,\overrightarrow{\varphi}_5(\tilde{z})
\end{equation}
where $d^a_{E,n}$ et $d^b_{E,n+1}$ are constants.

In order to determine the value of the constants ${a}^a_{E,n}$,
${a}^b_{E,n+1}$, $d^a_{E,n}$ and $d^b_{E,n+1}$ appearing in
(\ref{phirightmv}) and (\ref{phileftmv}), we have to solve the
system of equations formed by the continuity conditions of the wave
function and its first derivative at an intermediate point
$\tilde{z}_0\in[\tilde{z}_{\mathrm{min}},\tilde{z}_{\mathrm{max}}]$
\begin{equation}
\setlength{\arraycolsep}{2.5pt}
    \left\{\begin{array}{lll}
    \overrightarrow{\varphi}_{\mathrm{left}}(\tilde{z}_0)&=&\overrightarrow{\varphi}_{\mathrm{right}}(\tilde{z}_0)\vspace{5pt}\\
\overrightarrow{\varphi}^{\,\prime}_{\mathrm{left}}(\tilde{z}_0)&=&\overrightarrow{\varphi}^{\,\prime}_{\mathrm{right}}(\tilde{z}_0)
    \end{array}\right.
\end{equation}

Once this system has been solved, the value of the amplitude
$d^b_{E,n+1}$ is known and the induced emission probability can be
computed by use of Eq.~(\ref{Pem}).

\bibliographystyle{epj}
\bibliography{qobib}

\end{document}